\documentclass[preprint,11pt,natbib]{aastex}

\slugcomment{}

\shorttitle{ASYMMETRIC MAGNETIC RECONNECTION}
\shortauthors{MURPHY ET AL.}

\newcommand{\lya}{{\rm Ly}{$\alpha$}}
\newcommand{\lyb}{{\rm Ly}{$\beta$}}

\newcommand{\ciii}{{\rm C}~{\sc iii}}

\newcommand{\sixii}{{\rm Si}~{\sc xii}}

\newcommand{\ylt}{\ensuremath{y_{LT}}}

\newcommand{\cc}{cm\ensuremath{^{-3}}}

\newcommand{\SOHO}{\emph{SOHO}}

\newcommand{\Yohkoh}{\emph{Yohkoh}}
\newcommand{\Hinode}{\emph{Hinode}}
\newcommand{\RHESSI}{\emph{RHESSI}}

\newcommand{\e}{\ensuremath{\mathrm{e}}}
\newcommand{\dif}{\ensuremath{\mathrm{d}}}

\newcommand{\bhat}{\ensuremath{\mathbf{\hat{b}}}}
\newcommand{\xhat}{\ensuremath{\mathbf{\hat{x}}}}
\newcommand{\yhat}{\ensuremath{\mathbf{\hat{y}}}}
\newcommand{\zhat}{\ensuremath{\mathbf{\hat{z}}}}

\begin{document}

\title{ASYMMETRIC MAGNETIC RECONNECTION IN SOLAR FLARE AND CORONAL
  MASS EJECTION CURRENT SHEETS}

\author{N.~A.~Murphy,\altaffilmark{1}
        M.~P.~Miralles,\altaffilmark{1}
        C.~L.~Pope,\altaffilmark{1,2}
        J.~C.~Raymond,\altaffilmark{1}
        H.~D.~Winter,\altaffilmark{1} 
	K.~K.~Reeves,\altaffilmark{1}
	D.~B.~Seaton,\altaffilmark{3}
	A.~A.~{v}an~Ballegooijen,\altaffilmark{1}
	and 
        J.~Lin\altaffilmark{1,4}
}

\altaffiltext{1}{Harvard-Smithsonian Center for Astrophysics, 60
  Garden Street, Cambridge, MA 02138, USA}
\altaffiltext{2}{Elmhurst College, Elmhurst, IL, USA}
\altaffiltext{3}{SIDC-Royal Observatory of Belgium, Avenue Circulaire
  3, 1180 Brussels, Belgium} 
\altaffiltext{4}{Yunnan Astronomical Observatory, Chinese Academy of
  Sciences, P.O.\ Box 110, Kunming, Yunnan 650011, China}

\begin{abstract}
  We present two-dimensional resistive magnetohydrodynamic simulations
  of line-tied asymmetric magnetic reconnection in the context of
  solar flare and coronal mass ejection current sheets.  The
  reconnection process is made asymmetric along the inflow direction
  by allowing the initial upstream magnetic field strengths and
  densities to differ, and along the outflow direction by placing the
  initial perturbation near a conducting wall boundary that represents
  the photosphere.  When the upstream magnetic fields are asymmetric,
  the post-flare loop structure is distorted into a characteristic
  skewed candle flame shape.  The simulations can thus be used to
  provide constraints on the reconnection asymmetry in post-flare
  loops.  More hard X-ray emission is expected to occur at the
  footpoint on the weak magnetic field side because energetic
  particles are more likely to escape the magnetic mirror there than
  at the strong magnetic field footpoint.  The footpoint on the weak
  magnetic field side is predicted to move more quickly because of the
  requirement in two dimensions that equal amounts of flux must be
  reconnected from each upstream region.  The X-line drifts away from
  the conducting wall in all simulations with asymmetric outflow and
  into the strong magnetic field region during most of the simulations
  with asymmetric inflow.  There is net plasma flow across the X-line
  for both the inflow and outflow directions.  The reconnection
  exhaust directed away from the obstructing wall is significantly
  faster than the exhaust directed towards it.  The asymmetric inflow
  condition allows net vorticity in the rising outflow plasmoid which
  would appear as rolling motions about the flux rope axis.
\end{abstract}


\section{INTRODUCTION\label{introduction}}

Flux rope models of coronal mass ejections (CMEs) predict the
formation of an elongated current sheet in the wake behind the rising
plasmoid \citep[e.g.,][]{sturrock:1966, hirayama:1974,
kopp:pneuman:1976, lin:2000}.  Reconnection in these current sheets
increases the flux contained within the rising plasmoid and reduces
the amount of flux confining the plasmoid to low heights.
\citet{lin:2004} predict that these current sheets contribute
substantially to the mass budgets of CMEs such that the final
parameters for CME evolution for interplanetary propagation are not
set until the flux rope reaches several solar radii.  CME current
sheets may also play an important role in the CME energy budget
\citep{rakowski:2007, reeves:2010, murphy:retreat, murphy:cmeheat}.

In recent years, several features identified as current sheets have
been observed during and after CMEs \citep{ciaravella:2002, ko:2003,
sui:holman:2003, webb:2003, lin:2005, bemporad:2006, ciaravella:2008,
bemporad:2008, lin:2009, vrsnak:2009, sainthilaire:2009,
schettino:2010, liu:2010, savage:2010, savage:2011:nov3, landi:2010,
landi:2012, reeves:2011, patsourakos:2011}.  These features usually
appear as long-lasting bright streaks when viewed edge-on in white
light observations by the Large Angle and Spectrometric Coronagraph
\citep[LASCO,][]{brueckner:1995} on the \emph{Solar and Heliospheric
Observatory} (\SOHO).  Several events observed by LASCO were also
observed by the Ultraviolet Coronagraph Spectrometer
\citep[UVCS,][]{kohl:1995} on \SOHO\ in [{\rm Fe}~{\sc xviii}] and
\sixii\@.  LASCO images occasionally show large blobs
\citep[e.g.,][]{lin:2005} that are possibly the result of merging
plasmoids \citep{fermo:2010, uzdensky:2010, loureiro:2012}, although
the UVCS detection of strong \ciii\ emission in a few of these
features has indicated that this low ionization state plasma was not
processed through the current sheet.  While these blobs have been
interpreted as propagating at or near the Alfv\'en velocity, $V_A$,
simulations show that large magnetic islands often propagate at a
velocity $\sim$2--4 times slower \citep[e.g.,][]{shen:2011}.
Frequently, these current sheet features are observed to drift or tilt
with time, including during the `Cartwheel CME' \citep{savage:2010,
landi:2010, landi:2012} observed by the X-Ray Telescope (XRT) on
\Hinode\ on 2008 April 9\@.  The magnetic field topology and field
strengths in and around CME current sheets are not well understood
because of the lack of appropriate diagnostics.

Magnetic reconnection with asymmetry in the inflow direction occurs in
the Earth's dayside magnetopause \citep{phan:1996, ku:1997}, the
Earth's magnetotail \citep{oieroset:2004}, laboratory plasma
experiments \citep{yamada:1997, murphy:mrx}, plasma turbulence
\citep{servidio:2009, servidio:2010}, and during the merging of
unequal flux ropes \citep{linton:2006B}.  The scaling of asymmetric
inflow reconnection has been investigated by \citet{cassak:asym,
cassak:hall, cassak:dissipation} who find that the reconnection rate
depends on a hybrid Alfv\'en speed that is a function of the density
and magnetic field strength in both upstream regions, given in
dimensionless form by
\begin{equation}
  V_{Ah}^2 = \frac{B_LB_R(B_L+B_R)}{\rho_L B_R + \rho_R B_L},
  \label{cassakvah}
\end{equation}
where $B_L$ and $B_R$ are the upstream magnetic field strengths and
$\rho_L$ and $\rho_R$ are the upstream densities \citep[see
  also][]{borovsky:2007, birn:2008, birn:2010}.  Simulations
persistently show that the flow stagnation point and magnetic field
null are separated by a short distance, and that the X-line usually
drifts toward the region with the stronger magnetic field
\citep{ugai:2000, cassak:asym, murphy:mrx}.  The current sheet
drifting is less prevalent when a spatially localized resistivity is
imposed \citep[e.g.,][]{borovsky:2007, birn:2008}.  When there is a
pressure gradient along the inflow direction during guide field
reconnection, the X-line diamagnetically drifts along the outflow
direction.  Reconnection is suppressed when the drift velocity is
comparable to the Alfv\'en velocity \citep{rogers:1995, swisdak:2003,
  phan:2010, beidler:2011}.

CME current sheets form in a stratified atmosphere and thus are
expected to be asymmetric along the outflow direction as well.
Asymmetric outflow reconnection occurs in the solar atmosphere
\citep[][]{kopp:pneuman:1976, lin:2000, ciaravella:2008},
planetary magnetotails \citep[][]{oka:2011}, and laboratory
plasma experiments \citep{ono:1993, ono:1997, inomoto:2006,
2008PPCF...50g4012L, murphy:mrx, gray:2010}.  \citet{murphy:asym}
derived scaling relations for a long and thin current sheet with
asymmetric downstream pressure and found that the reconnection rate is
not greatly affected unless outflow is blocked from both ends of the
current sheet.  Simulations of asymmetric outflow reconnection and
X-line retreat show that the X-line and flow stagnation point are
separated by a short distance and that most of the energy released
during reconnection is directed towards the unobstructed current sheet
exit \citep{roussev:2001A, galsgaard:2002, oka:2008, reeves:2010,
murphy:retreat, shen:2011}.  Asymmetry and motion of magnetic nulls
have recently been considered in three-dimensional configurations
\citep[e.g.,][]{alhachami:2010, galsgaard:2011, gray:2010,
lukin:2011}.

In this work we simulate magnetic reconnection that is asymmetric in
both the inflow and outflow directions and consider the effects that
these asymmetries may have on solar flare and CME current sheets.
In Section \ref{method}, we discuss the numerical method used by the
NIMROD code and the simulation setup.
In Section \ref{results}, we present the simulation results including
the magnetic structure, flow pattern, X-line dynamics, vorticity in
the outflow plasmoid, and the morphology of the post-flare loops.
In Section \ref{analytic}, we derive an analytic solution for the
potential field structure of asymmetric post-flare loops and discuss
the connection with the simulation results.
In Section \ref{observational}, we discuss expected observational
signatures of line-tied asymmetric reconnection during solar
eruptions.
Section \ref{discussion} contains a summary and a discussion of our
results.

\section{NUMERICAL METHOD AND SIMULATION SETUP\label{method}}

The NIMROD code \citep{sovinec:jcp,sovinec:jop,sovinec:2010} solves
the equations of extended magnetohydrodynamics (MHD) using a finite
element formulation for the poloidal plane, and for three-dimensional
simulations, a finite Fourier series expansion for the out-of-plane
direction.  In dimensionless form, the equations solved for the
two-dimensional simulations reported in this paper are
\begin{eqnarray}
  \frac{\partial \rho}{\partial t}
  + \nabla \cdot \left( \rho \mathbf{V} \right)
  = \nabla \cdot D \nabla \rho,  \label{continuity}
  \\
  \frac{\partial \mathbf{B}}{\partial t} 
  =
  - \nabla \times 
  \left(
    \eta \mathbf{J} - \mathbf{V}\times\mathbf{B}
  \right), \label{farohms}
  \\
  \mathbf{J} = \nabla \times \mathbf{B}, \label{ampere}
  \\
  \rho 
  \left(
    \frac{\partial \mathbf{V}}{\partial t}
    + \mathbf{V} \cdot \nabla \mathbf{V}
  \right)
  = 
  \mathbf{J}\times\mathbf{B}
  - \nabla{p}
  + \nabla \cdot \rho \nu \nabla \mathbf{V}, \label{momentum}
  \\
  \frac{\rho}{\gamma-1}
  \left(
    \frac{\partial T}{\partial t} + \mathbf{V} \cdot \nabla T
  \right)
  =
  - \frac{p}{2} \nabla \cdot \mathbf{V}
  - \nabla \cdot \mathbf{q}
  + Q, \label{temperature}
\end{eqnarray}
where $\mathbf{B}$ is the magnetic field, $\mathbf{J}$ is the current
density, $\mathbf{V}$ is the bulk plasma velocity, $p$ is the plasma
pressure, $\rho$ is the density, $\eta$ is the resistivity, $\nu$ is
the kinematic viscosity, $D$ is an artificial number density
diffusivity, $T$ is the temperature, $\gamma=5/3$ is the ratio of
specific heats, and $Q$ includes Ohmic and viscous heating. The heat
flux vector includes the effects of anisotropic thermal conduction and
is given by $\mathbf{q}=-\rho \left[ \chi_\|\bhat\bhat +
  \chi_\perp\left( \mathbf{I} - \bhat\bhat \right) \right]\cdot\nabla
T$, where \bhat\ is a unit vector in the direction of the magnetic
field.
The normalizations are given by: $B_0$, $\rho_0$, $L_0$, $t_0$,
$V_{A0} \equiv B_0/\sqrt{\mu_0\rho} \equiv L_0/t_0$, $p_0\equiv
B_0/\mu_0 \equiv \rho_0 V_{A0}^2$, $J_0\equiv B_0/\mu_0L_0$, and
$\eta_0/\mu_0 \equiv \nu_0 \equiv \chi_0 \equiv D_0 \equiv L_0^2/t_0$.
Divergence cleaning is used to prevent the buildup of divergence error
\citep{sovinec:jcp}.
A small number density diffusivity ($D \ll \eta$) is included to
ensure that the number density profile remains sufficiently smooth in
regions of sharp gradients.
Except as noted, all velocities are in the simulation reference frame.
Additional details for the numerical method, normalizations, and
simulation setup are presented by \citet{murphy:retreat}.

Rather than modeling a solar eruption in detail, we choose an
idealized initial condition that allows asymmetric reconnection to
commence.  The initial conditions are of a perturbed, modified Harris
sheet that allows asymmetric upstream magnetic field strengths,
densities, and plasma pressure \citep[see][]{birn:2008, birn:2010}.
We define $\xhat$ as the inflow direction, $\yhat$ as the outflow
direction, and $\zhat$ as the out-of-plane direction.  The initial
equilibrium is given by
\begin{eqnarray}
  \mathbf{B}(x) =B_{R0}
  \left[
    \frac{\tanh \left(\frac{x}{\delta_0}-b\right)+b}{1+b}
  \right] \yhat + B_{z0}\,\zhat
  \\
  p(x) = \frac{1}{2}\left(1-B_y^2\right) +
    \beta_{R0}\frac{B_{R0}^2}{2}
  \\
  \rho(x) = \rho_0 \left[ 1 +
  \left(f_0-1\right)\left(1-\lambda^2\right)\right], 
  \\
    \lambda \equiv \frac{1}{2}\left[ 
    1 + \tanh\left( \frac{x}{\delta_0}-b\right)
  \right],
\end{eqnarray}
where $\delta_0$ is the initial current sheet thickness, $b$ is the
magnetic asymmetry parameter, $B_{z0}$ is the initial guide field
(equal to zero except in case F where $B_{z0}=4$), and $\beta_{R0} =
p_{R0} / \left(B_{R0}^2/2 \right)$.  Note that $\beta_{R0}$ does not
include the guide field contribution.
The subscripts `$L$' and `$R$' refer to the asymptotic
magnitudes of quantities for $x<0$ and $x>0$, respectively, and the
subscript `$0$' refers to the values of quantities at the beginning of
each simulation.  The initial ratios for the upstream densities and
magnetic fields are given by
\begin{eqnarray}
  f_0 & \equiv & \frac{\rho_{L0}}{\rho_{R0}},   \label{fdef}
  \\
  R_0 & \equiv & \frac{B_{L0}}{B_{R0}} = \frac{1-b}{1+b}, \label{rdef}
\end{eqnarray}
where $0 \leq b < 1$, $B_{L0}>0$, and $B_{R0}>0$.  These ratios at
$t=0$ will in general differ somewhat from the ratios during the
course of each simulation.

The initial magnetic perturbation is of the form
\begin{equation}
  \mathbf{B}_p = 
  \nabla\times\left(A_p\zhat\right),
\end{equation}
where
\begin{equation}
  A_p = -B_p h
  \exp
  \left[
    -\left(\frac{x}{h}\right)^2
    -\left(\frac{y-\Delta}{h}\right)^2
  \right].
\end{equation} 
Here, $B_p$ is the strength of the perturbation and $h$ governs the
perturbation width. The perturbation is centered about
$(x,y)=(0,\Delta)$.

The simulations we report on are presented in Table \ref{simtable}\@.
The controls for our study are case A, with symmetric inflow and
asymmetric outflow, and case H, with asymmetric inflow and symmetric
outflow.  The eight simulations in Table \ref{simtable} are chosen to
test different upstream magnetic field asymmetries (compare cases
A--D), different resistivities (cases C and E), the inclusion of a
guide field (cases C and F), and different upstream densities (cases A
and G).  Case C is a control for comparisons with cases E, F, and H
which all have $R_0=0.25$\@.

The simulation parameters are as follows.  For the initial equilibrium
we use $B_0=1$, $\delta_0=0.1$, and $\beta_{R0}=0.18$.
The initial perturbation is given by $B_p=0.1$, $\Delta=1$, and
$h=0.5$, except for case H where $\Delta=0$.
The diffusivities are given by $\eta = \nu = 10^{-3}$, $D= 10^{-4}$,
$\chi_\perp = 10^{-4}$, and $\chi_\| = 10^{-2}$.  The Lundquist number
must be considered carefully because the Alfv\'en speeds differ in
each upstream region.  For our comparisons we use a hybrid Lundquist
number based on the hybrid Alfv\'en speed presented in Equation
(\ref{cassakvah}),
\begin{equation}
S_h \equiv \frac{LV_{Ah}}{\eta},\label{lundquist}
\end{equation}
where $L$ is the characteristic length scale of the current sheet.
Mesh packing is used to concentrate resolution in regions of strong
gradients.  However, because the current sheets drift in most
simulations, high resolution is needed over a larger part of the
domain than in simulations with symmetric inflow and/or outflow.
We use seventh order finite element basis functions in all
simulations.  There are $m_x$ and $m_y$ finite elements along the
inflow and outflow directions, respectively.
The size of the computational domain for cases A--G is given by $-7
\leq x \leq 7$ and $0 \leq y \leq 30$.  
We assume conducting wall outer boundary conditions in all directions
for cases A--G; consequently, late in time there is some influence
from line-tying along the upper boundary at $y=30$.  However, we
concentrate on dynamics far from the unphysical upper boundary ($y
\lesssim 18$).
Case H differs in that $-15 \leq y \leq 15$ and there are periodic
boundary conditions in the $y$-direction. 
While we include anisotropic thermal conduction, we neglect radiative
losses, coronal heating, and vertical stratification of the
atmosphere.  The cooling time scale for CME current sheet plasma ($T
\sim 5\times 10^6$~K and $n\sim 5\times 10^8$~\cc) is about half a
day.  This is an order of magnitude longer than a dynamical time scale
so we are justified in neglecting radiative losses \citep[see
also][]{imada:2011:nei}.

It is appropriate to remark upon our choice of a uniform, explicitly
defined resistivity.  Occasionally, prior simulations of asymmetric
inflow reconnection have used a spatially localized resistivity
enhancement \citep[e.g.,][]{borovsky:2007}.  The intended
effect is to constrain the position of the X-line.  However, this
constraint is artificial and likely changes the internal structure of
the reconnection region and the scaling behavior.  We speculate that
this is the reason why the scaling found by \citet{borovsky:2007}
differs slightly from the analytic prediction made by
\citet{cassak:asym}.
Some simulations use resistivity as a function of current density or
other plasma parameters so that the position of the X-line is not
artificially constrained \citep[see, for example,][]{birn:1996B}.
Alternatively, some simulations of reconnection have used numerical
resistivity inherent in the discretization as the sole means of
breaking the frozen-in condition, rather than an explicitly defined
resistivity or field line breaking mechanism included in the physical
model \citep[e.g.,][]{laitinen:2005, ouellette:2010, edmondson:2010}.
Such models are effectively using an ideal MHD algorithm.  However,
when there is significant numerical reconnection in an ideal MHD
simulation, the results are by definition not converged.  Therefore,
while scaling models for asymmetric reconnection such as those by
\citet{cassak:asym} and \citet{murphy:asym} do not explicitly depend
on the dissipation mechanism, the lack of convergence makes ideal MHD
simulations poor tests of asymmetric reconnection models.
Our choice of a uniform, explicitly defined resistivity avoids
artificially constraining the position of the X-line and allows
convergence in the numerical simulations.

\section{SIMULATION RESULTS\label{results}}
In this section, we describe the principal results of our simulations
of line-tied asymmetric reconnection.  When making comparisons between
the simulations presented in Table \ref{simtable} it is important to
note that the scaling for cases A--D is performed by keeping $B_{R0}$
constant and reducing $B_{L0}$.  Consequently, the total magnetic
energy available to be reconnected and the hybrid Alfv\'en speed given
in Equation (\ref{cassakvah}) decrease as the simulations become more
asymmetric.  Case G is best compared directly to case A, and cases E,
F, and H should be compared to case C\@.

\subsection{Principal Features}

\begin{figure}
  \begin{center}
    \includegraphics[scale=0.9]{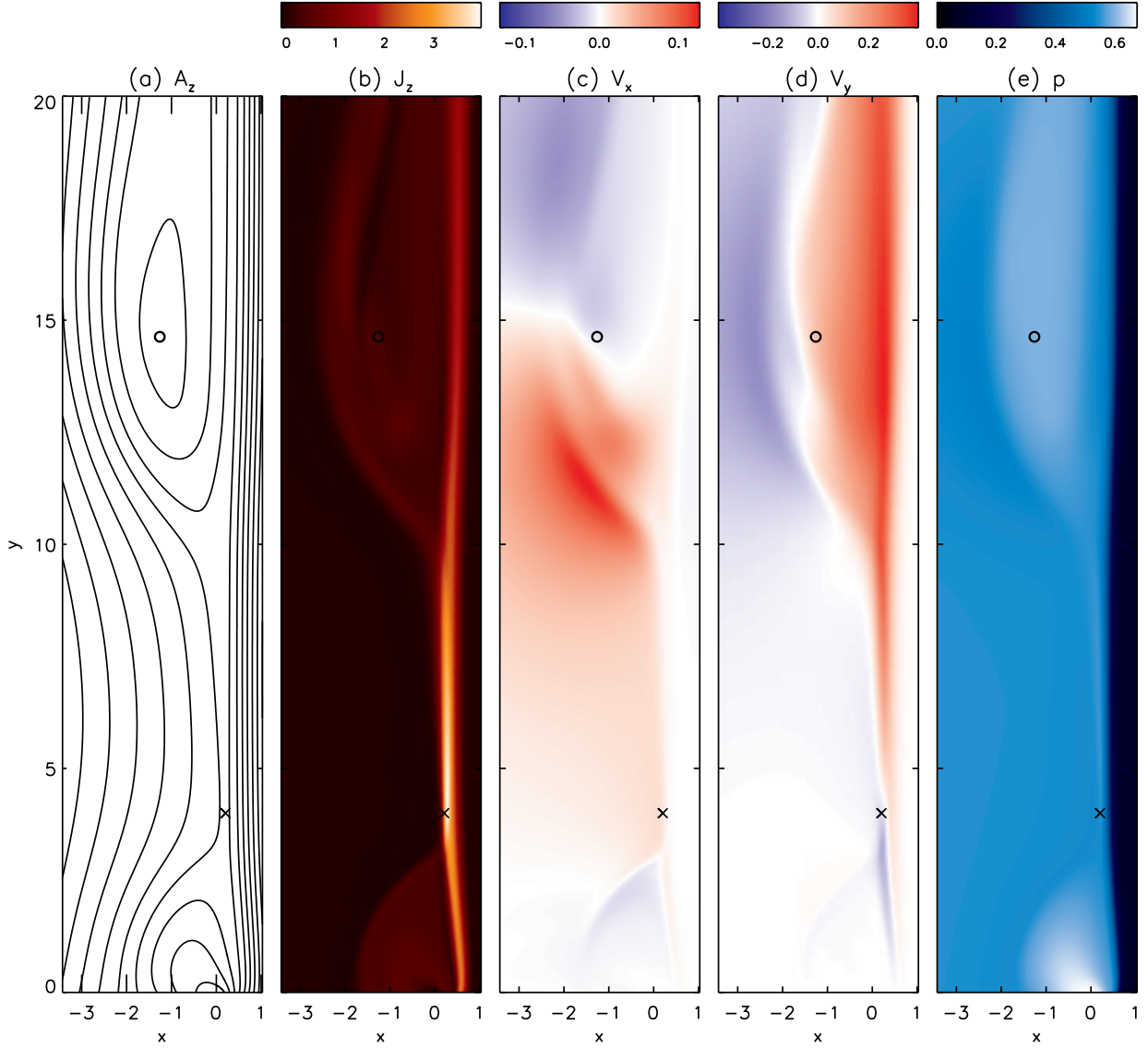}
  \end{center}
  \caption{Simulation results for case C with $R_0=B_{L0}/B_{R0}=0.25$
    at $t=100$.  Shown are (a) the magnetic flux, $A_z$, (b) the
    out-of-plane current density, $J_z$, (c) the inflow component of
    velocity, $V_x$, (d) the outflow component of velocity, $V_y$, and
    (e) the plasma pressure, $p$.  The `$\circ$' denotes the position
    of the O-point, and the `{$\times$}' marks the spot of the X-line.
    Only a portion of the computational domain is shown.
  \label{paperplot_simulation}}
\end{figure}

The general features of our simulations of line-tied asymmetric
reconnection are presented in Figure \ref{paperplot_simulation}, which
shows case C with an initial upstream magnetic field ratio of
$R_0=0.25$.  Magnetic flux contours in Figure
\ref{paperplot_simulation}(a) show that field lines in the strong
magnetic field region are much less bent than field lines in the weak
magnetic field region.  The reconnected loops near the lower boundary
and the outflow plasmoid both preferentially develop into the weak
magnetic field region.  The separatrices are traced by regions of
strong out-of-plane current density in Figure
\ref{paperplot_simulation}(b), but the portions that bound the strong
field regions have much stronger current density than the portions
that form the boundary between the outflow and weak field regions.
The plasma pressure buildup is almost entirely contained on the weak
field side of the current sheet [Figure
  \ref{paperplot_simulation}(e)].

The upward outflow velocity, shown in Figure
\ref{paperplot_simulation}(d), is substantially faster than the
downward outflow velocity because of the obstructing wall along $y=0$.
This behavior is consistent with previous simulations of asymmetric
outflow reconnection \citep{roussev:2001A, galsgaard:2002, oka:2008,
murphy:mrx, murphy:retreat, reeves:2010}.  This difference in
velocities occurs for two principal reasons.  First, outflow towards
$y=0$ is obstructed by the buildup of plasma and magnetic pressure.
Second, the X-line is located near the lower exit of the current sheet
so that the tension force below the X-line directed towards $y=0$ is
much weaker than the tension force above the X-line directed away from
$y=0$.

In Figure \ref{paperplot_simulation}(d), there is a stream of plasma
flow to the right of the X-line but extending downward into the
post-flare loop structure and tracing the separatrix.  The magnitude
of the plasma flow is small ($V_y\sim 0.02$ compared to
$V_{Ah0}=0.25$), but is positive in a region where negative $V_y$ is
expected.  Magnetic tension in this region is pulling plasma downward,
but is countered by a comparable contribution by the vertical plasma
pressure gradient and a modest contribution by the vertical magnetic
pressure gradient pushing plasma upward.  The current density is
strong so that resistive diffusion acts to make the field lines more
potential even against this plasma flow.  Such flows might be
observable, but three-dimensional geometry may make identification
ambiguous, and it is not clear that they will occur when $\beta\ll 1$.

\subsection{Internal Structure\label{internal}}

\begin{figure}
  \begin{center}
    \includegraphics[scale=1.0]{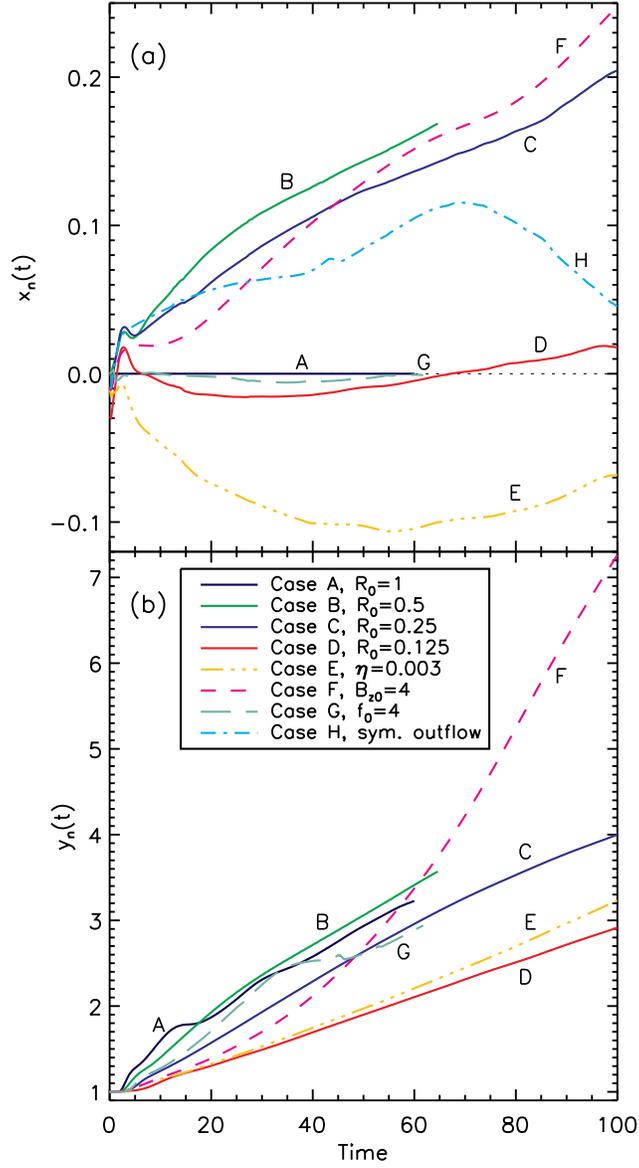}
  \end{center}
  \caption{The position of the X-line as a function of time for cases
    A--H along the (a) inflow and (b) outflow directions.  Case H is
    not shown in the bottom panel because $y_n(t)=0$ due to symmetry.
  \label{paperplot_xn}}
\end{figure}

The X-line (the line about which the magnetic field has a
hyperbolic or X-line topology) is located at $\textbf{x}_n\equiv
\left(x_n,y_n\right)$.  This position
is given as a function of time for several simulations in Figure
\ref{paperplot_xn}.  Figure \ref{paperplot_xn}(a) shows the X-line
position along the inflow direction.  In most previous simulations of
asymmetric inflow reconnection, the X-line drifts into the region with
the stronger magnetic field \citep[e.g.,][]{ugai:2000}.  This drifting
is tied locally to spatial derivatives in the out-of-plane electric
field \citep{murphy:retreat}.  In particular, the X-line drifts along
the inflow direction towards decreasing $E_z$.  For many but not all
of the cases we observe the X-line drifting towards the strong field
upstream region for most of the simulation.  
In case D with $R_0=0.125$, the X-line drifts into the weak field
region before drifting into the strong field region.  Case E is
identical to case C except for having three times the resistivity, and
shows a significant drift of the X-line into the weak magnetic field
region.  Comparing cases A and G suggests that the inclusion of a
density asymmetry does not lead to a significant drift along the
inflow direction.
The rate of X-line drifting along the inflow direction is a function
of the magnetic field asymmetry, the resistivity, the inclusion of a
guide field, and the simulation setup.  However, the dependences on
each of these parameters are not straightforward.

The X-line position retreats from the obstruction at $y=0$ for all
asymmetric outflow cases [Figure \ref{paperplot_xn}(b)].  The rate of
X-line retreat is comparable for cases A and B (with $R_0=1$ and
$R_0=0.5$, respectively) but decreases as the initial conditions
become more asymmetric.
By comparing cases C and E we see that increasing the resistivity
actually slows down X-line retreat, in part because it results in
smoother gradients in $E_z$.
From cases A and G we see that increasing $f_0$ from 1 to 4 decreases
the rate of X-line retreat slightly.  
Comparing cases C and F shows that including a guide field can lead to
much faster X-line retreat.  In contrast to all other asymmetric
outflow simulations, the X-line is located near the top exit of the
current sheet in case F.  

Previous simulations of reconnection with either asymmetric inflow or
asymmetric outflow have generally shown a separation between the flow
stagnation point and the principal X-line when a uniform resistivity
is used \citep[e.g.,][]{cassak:asym, cassak:hall, cassak:dissipation,
  oka:2008, murphy:mrx, murphy:retreat, shen:2011}.  Contrary to our
expectations, there is in general no flow stagnation point associated
with the X-line in the simulation reference frame for cases with both
asymmetric upstream magnetic fields and asymmetric outflow.
There are contours where $V_x=0$ and $V_y=0$, but these contours
do not intersect with each other in the current sheet for most of the
simulations with asymmetry in both the inflow and outflow directions.
There still is a reversal of the outflow component of velocity near
but not colocated with the X-line.

\begin{figure}
  \begin{center}
    \includegraphics[scale=1.0]{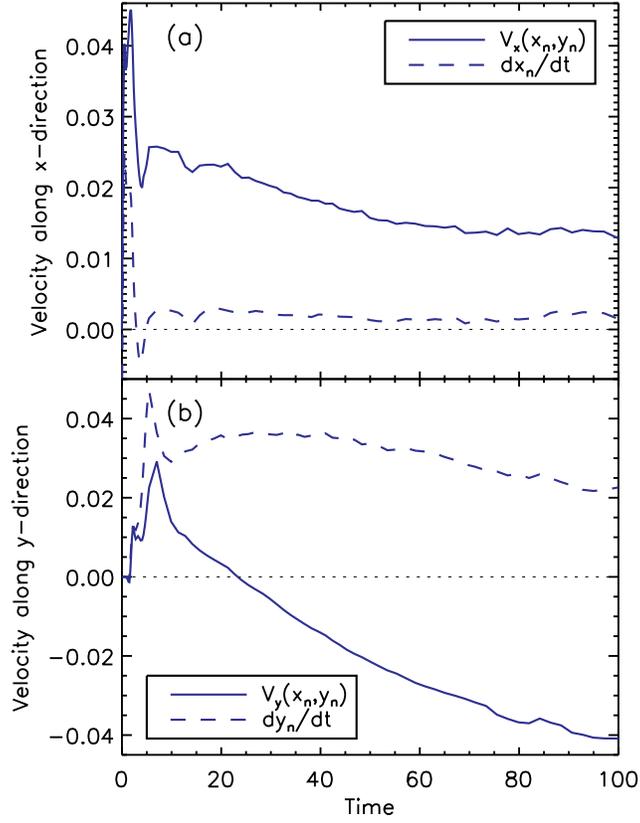}
  \end{center}
  \caption{A comparison of the plasma flow velocity at the X-line,
    $(V_x(x_n,y_n),V_y(x_n,y_n))$, to the time derivative of the
    position of the X-line, $(\dif x_n/\dif t,\dif y_n/\dif t)$, for
    the (a) inflow direction and (b) outflow direction, for case C
    with $R_0=0.25$.  The plasma flow velocity at the X-line differs
    significantly from the rate of X-line motion along both
    directions.
  \label{paperplot_derivs}}
\end{figure}

Figure \ref{paperplot_derivs} shows both the plasma flow velocity at
the X-line, $\mathbf{V}_n\equiv \left(V_x\left(x_n,y_n\right),
V_y\left(x_n,y_n\right)\right)$, and the rate of change in position of
the X-line, $\dot{\mathbf{x}}_n\equiv \left(\dot{x}_n,
\dot{y_n}\right)$, for case C with $R_0=0.25$. The plasma flow
velocity at the X-line differs greatly from the X-line drift velocity
along both the inflow and outflow directions, indicating a significant
departure from the frozen-in condition.  For the inflow direction,
$V_x(x_n,y_n)$ is of the same sign as $\dot{x}_n$, but $V_x(x_n,y_n)$
is significantly greater.  For the outflow direction, $\dot{y}_n$
remains positive but $V_y(x_n,y_n)$ becomes negative so that the
X-line is retreating against the flow of the plasma.  The X-line is
able to diffuse against strong plasma flow by diffusion of $B_x$ along
the inflow direction, as described in detail by
\citet{murphy:retreat} \citep[see also][]{siscoe:2002, oka:2008}.

\begin{figure}
  \begin{center}
    \includegraphics[scale=1.0]{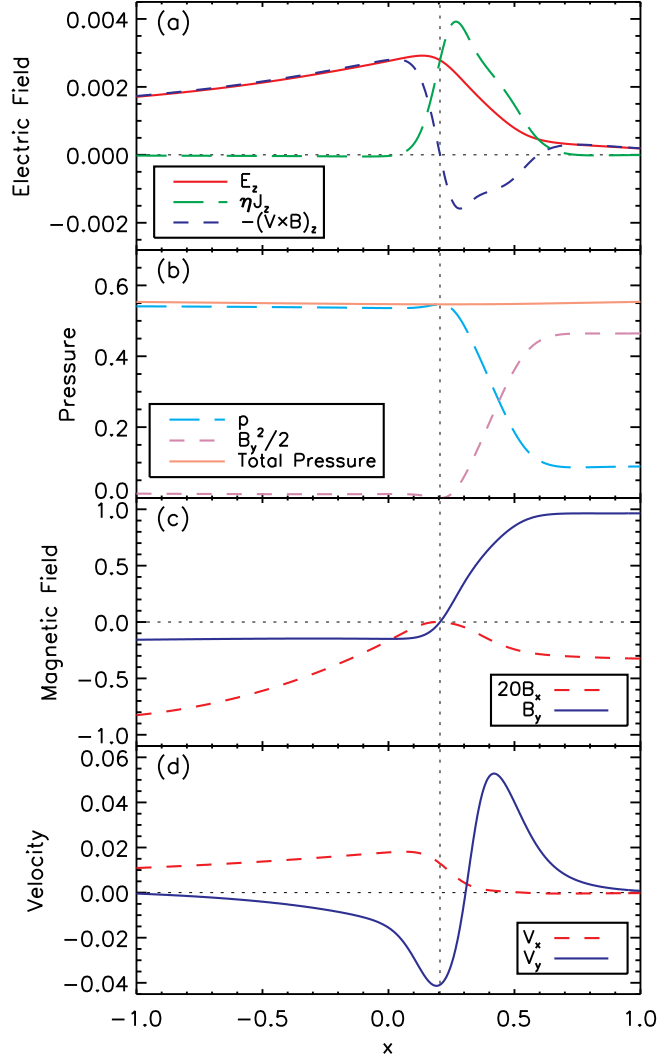}
  \end{center}
  \caption{ Simulation parameters for a slice along the inflow
    direction at the position of the X-line for case C with $R_0=0.25$
    at $t=100$.  The vertical dotted line represents the position of
    the X-line along the inflow direction.  Shown are (a) components
    of the out-of-plane electric field, (b) contributions to total
    pressure balance along the inflow direction, (c) components of the
    in-plane magnetic field, and (d) the inflow and outflow components
    of velocity.
    \label{paperplot_inflow}
  }
\end{figure}

A slice along the inflow direction for case C at the height of the
X-line ($y=y_n$) is given in Figure \ref{paperplot_inflow} for
$t=100$.
The maximum in $J_z$ is located on the strong field side of the
X-line.  This result that the X-line is on the weak field side of the
current sheet is consistent with the simulations of
\citet{cassak:asym}.
The X-line is located near a local maximum in plasma pressure along
the inflow direction.
In the simulation reference frame, $V_x\approx 0$ away from the
diffusion region on the strong field side of the current sheet.  This
can be interpreted as most of the inflow coming in from the weak field
side, with the X-line and current sheet drifting into the strong field
upstream region.
The outflow component of velocity, $V_y$, has an interesting bipolar
signature.  While the X-line is moving in the positive $y$ direction,
the X-line is located near where $V_y$ is most negative.
The inflow component of the magnetic field, $B_x$, is negative
throughout the slice except at the X-line where it is zero.  This
profile allows negative $B_x$ to diffuse inward so that at later times
the X-line is located at higher heights \citep[compare to Figure 5
  of][]{murphy:retreat}.
The reconnecting component of the magnetic field, $B_y$, shows that
the initial ratio of $R_0=0.25$ is somewhat less extreme than the
ratio of $R\approx 0.17$ at $t=100$.

\subsection{Reconnection Rate\label{reconrate}}

\begin{figure}
  \begin{center}
    \includegraphics[scale=1.0]{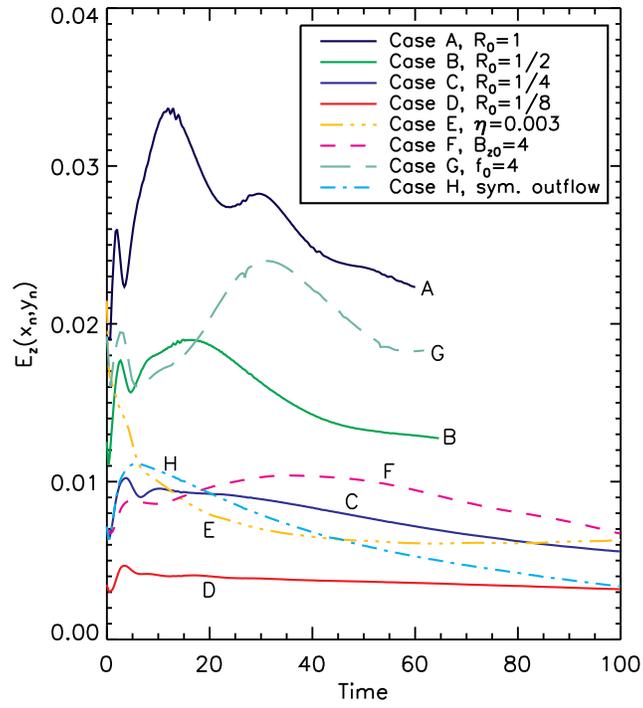}
  \end{center}
  \caption{Reconnection electric field strengths at the X-line as a
    function of time for cases A--H\@.
  \label{paperplot_ez}}
\end{figure}

The reconnection rate, defined as the out-of-plane component of the
electric field at the X-line, is shown in Figure \ref{paperplot_ez}
for each of the cases described in Table \ref{simtable}.  Comparing
cases A--D shows that decreasing the magnetic asymmetry factor $R_0$
leads to a corresponding decrease in the reconnection rate. This is
qualitatively consistent with the scaling derived by
\citet{cassak:asym}.  
Increasing the density asymmetry factor $f_0$ decreases the
reconnection rate modestly from case A to case G\@.
The reconnection rate in case C is somewhat quicker than case E, even
though case E is three times more resistive.
The inclusion of a guide field slightly increases the reconnection
rate in case F compared to case C\@.  
The reconnection rate is modestly quicker with asymmetric inflow and
outflow (case C) than in an otherwise equivalent simulation with
asymmetric inflow but symmetric outflow.  As described by
\citet{murphy:retreat}, this occurs because the current sheet can only
increase in length along one outflow direction.

\subsection{The Outflow Plasmoid\label{plasmoid}} 

\begin{figure}
  \begin{center}
  \includegraphics[scale=0.8]{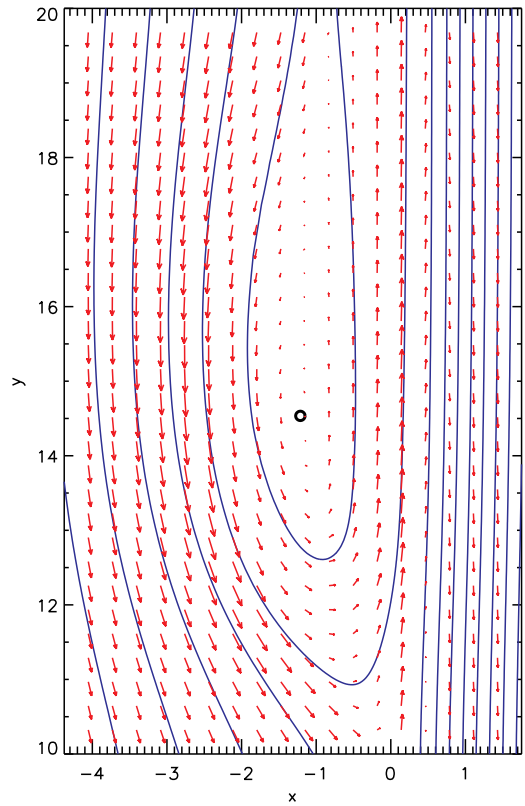}
  \end{center}
  \caption{ Shown are the magnetic flux (solid blue contours) and
    flow velocity vectors in the frame of the O-point (red arrows) for
    case C at $t=100$.  The O-point is denoted by the circle at
    $(x_\mathrm{o},y_\mathrm{o}) = (-1.21,14.53)$.  The flow pattern
    about the O-point is counterclockwise; consequently there is net
    vorticity in the outflow plasmoid.
  \label{paperplot_vorticity}}
\end{figure}

The asymmetric inflow condition allows net vorticity in the outflow
plasmoid.  In Figure \ref{paperplot_vorticity} we show velocity
vectors in the reference frame of the O-point for case C at
$t=100$. The counter-clockwise flow pattern is largely due to the
reconnection outflow jet impacting the rising plasmoid at an angle;
consequently, the outflow jet increases the net vorticity in the
plasmoid.  This circulation should be a generic feature of asymmetric
inflow reconnection but not asymmetric outflow reconnection.  This
interpretation should be considered qualitatively rather than
quantitatively because the presence of closed field lines is partially
due to reconnection near the outer, artificial line-tied boundary at
$y=30$.  However, these results suggest that the orientation of
reconnection outflow jets relative to the flux rope can lead to an
analogous circulation pattern when the outflow jet impacts the flux
rope obliquely rather than directly at its base.


In contrast to the X-point, the frozen-in condition near the O-point
is approximately met.  For example, at $t=100$ in case C, the plasma
flow at the O-point is approximately equal to the velocity of the
O-point: $(V_x(x_o,y_o),V_y(x_o,y_o)) = (-0.0076,0.1211)$ compared to
$(\dot{x}_o,\dot{y}_o) = (-0.0065,0.1238)$, where $\mathbf{x}_o \equiv
(x_o,y_o)$ is the position of the O-point.  This indicates that the
O-point is being primarily advected by the bulk plasma flow and that
diffusive flow across the O-point is not significant.  Plasma flow
across an O-point can occur in a process similar to that described by
\citet{murphy:retreat} for X-line retreat. When the O-point in a
magnetic island is displaced towards one particular direction,
resistive diffusion will in general act to change the position of the
O-point to be closer to the center of the island when the island is
not significantly distorted.

\subsection{Morphology of the Post-Flare Loops\label{loops}}

The most easily observable difference between the symmetric and
asymmetric cases is the structure of the post-flare loops.  In
particular, the asymmetric loops are skewed when compared to the
symmetric case, and take an asymmetric candle flame shape.  While in
the symmetric case the tops of each loop are all along $x=0$, the
loop-tops in the asymmetric case have their apexes at different
locations along the inflow direction (Figure \ref{paperplot_loops}).
At low heights where the field has relaxed to a near-potential state,
the loop-tops are displaced toward the low magnetic field side.  At
greater heights, the loop-tops become located closer to the current
sheet demarcating the low and high magnetic field regions.

\section{AN ANALYTIC SOLUTION FOR THE POTENTIAL FIELD STRUCTURE OF
  ASYMMETRIC POST-FLARE LOOPS\label{analytic}}

\begin{figure}
  \begin{center}
  \includegraphics[scale=1.0]{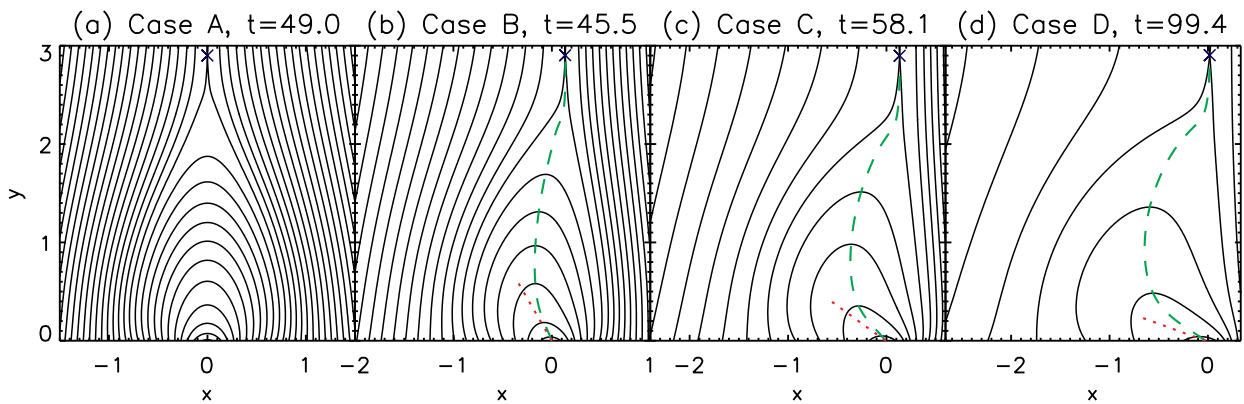}
  \caption{Magnetic flux contours for cases A--D when the X-line in
    each simulation (denoted by `$\times$') is located at $y=2.9$.
    The dashed green line gives the position of the loop apexes (where
    $B_y=0$).  The red dotted line represents the asymptotic
    prediction made by Equation (\ref{magic}).
    \label{paperplot_loops}}
  \end{center}
\end{figure}

In Section \ref{loops} and Figure \ref{paperplot_loops} we show that
the location of the post-flare loop apexes in each case are a function
of height.  In the current section, we present an analytic solution
for the potential field structure of post-flare loops in the present
configuration.  This solution provides insight into the observational
signatures very near the magnetic field reversal along the lower
boundary, but should be treated as a limiting case for post-flare
loops late in time that have been able to relax to a near-potential
state.

We consider the domain $0\leq x \leq \pi$ and $y \geq 0$.  The
boundary condition along $y=0$ is given by
\begin{equation}
  B_y(x,0) =
  \left\{ 
  \begin{array}{r@{\quad:\quad}l}
    -B_L & 0 < x < a \\
    B_R & a<x<\pi
  \end{array}
  \right., \label{lowerbc}
\end{equation}
where the constants $B_L$ and $B_R$ are positive and the location of
the field reversal is given by
\begin{equation}
a = \frac{\pi B_R}{B_L+B_R}.
\end{equation}
Consequently, there is no net magnetic flux from the lower boundary
over $0\leq x \leq \pi$.  Because the outer boundary is artificial, we
must consider regions close to the magnetic field reversal to avoid
these effects.  However, this analysis also assumes that the magnetic
field reversal length scale is smaller than the region being
investigated.  Because we are finding a potential field solution we
also require that $\beta \ll 1$, or more generally, that plasma
pressure gradient forces are small.

\begin{figure}
  \begin{center}
  \includegraphics[scale=1.0]{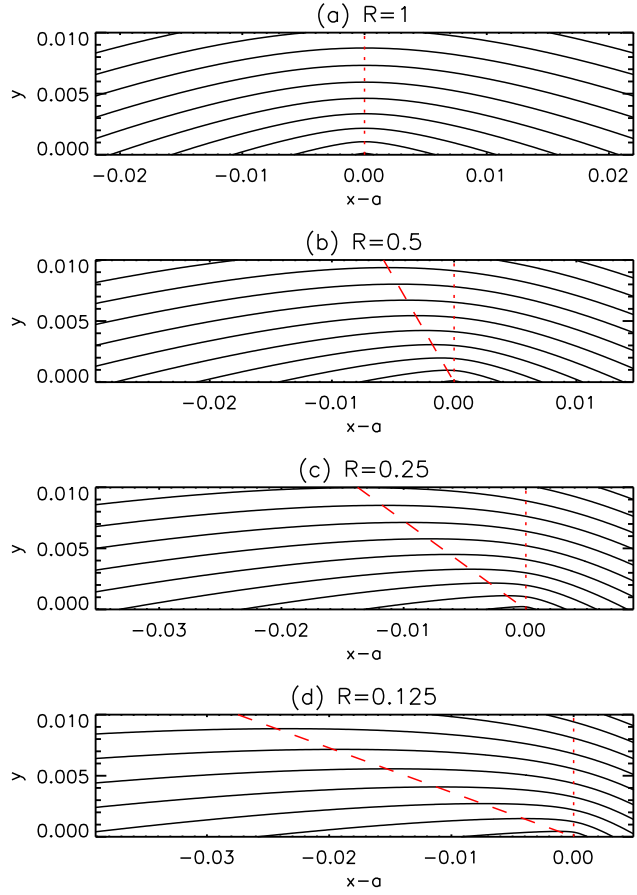}
  \end{center}
  \caption{Magnetic flux contours for the analytical potential field
    solution of asymmetric post-flare loops as given by Equation
    (\ref{vpsolution}) very near the magnetic field reversal along the
    lower boundary at $x=a$ (represented by the red dotted vertical
    line).  The dashed red line is the location of the loop-tops as
    given by Equation (\ref{magic}).  The solutions are for magnetic
    field ratios of (a) $R=1$, (b) $R=0.5$, (c) $R=0.25$, and (d)
    $R=0.125$.
    \label{analytic_solution}
}
\end{figure}

The potential magnetic field is given by
\begin{equation}
  \mathbf{B} = - \nabla\psi,\label{potential}
\end{equation}
where the scalar potential $\psi$ is governed by Laplace's equation,
\begin{equation}
  \nabla^2 \psi = 0.
\end{equation}
The unique solution to Laplace's equation appropriate to our boundary
conditions is
\begin{equation}
  \psi = 
  \sum_{n=1}^\infty 
  \frac{2}{\pi}
  \left(\frac{B_L+B_R}{n^2}  \right)
  \sin{na} \cos{nx}\;\e^{-ny}.
  \label{solution}
\end{equation}
Equation (\ref{solution}) may be written in the form of a vector
potential $\mathbf{B}=\nabla\times\left(A_z\zhat\right)$ to provide an
expression for the magnetic flux,
\begin{eqnarray}
   A_z = 
   \sum_{n=1}^\infty 
   \frac{2}{\pi}
   \left(\frac{B_L+B_R}{n^2}  \right)
   \sin{na} \sin{nx}\;\e^{-ny}.
  \label{vpsolution}
\end{eqnarray}
On large scales, the solution depends on the outer boundary.  On
scales much smaller than the outer boundary, the solution becomes
scale-free as we approach $(x,y)\rightarrow (a,0)$.  Because in
general Fourier series expansions will be truncated at some
$N_{\mathrm{max}}$, we must also consider scales $\Delta x \gg
\pi/N_{\mathrm{max}}$.  Using Euler's formula and the identity
$\ln\left(1-q\right) = -\sum_{n=1}^\infty q^n/n$, the components of
the magnetic field are given  in closed form by
\begin{eqnarray}
  B_x &=& \frac{B_L+B_R}{2\pi}
  \ln
  \left[
    \frac{
      \left( 1 - \e^{ ia+ix-y} \right)
      \left( 1 - \e^{-ia-ix-y} \right)
    }{
      \left( 1 - \e^{ ia-ix-y} \right)
      \left( 1 - \e^{-ia+ix-y} \right)
    }
  \right],\label{bxsolution}
  \\
  B_y &=& i\left(\frac{B_L+B_R}{2\pi}\right)
  \ln
  \left[
    \frac{
      \left( 1 - \e^{ ia+ix-y} \right)
      \left( 1 - \e^{ ia-ix-y} \right)
    }{
      \left( 1 - \e^{-ia+ix-y} \right)
      \left( 1 - \e^{-ia-ix-y} \right)
    }
  \right].\label{bysolution}
\end{eqnarray}
Solutions for four different magnetic field ratios are presented in
Figure \ref{analytic_solution} very close to the magnetic field
reversal along the lower boundary.  These solutions approximate the
magnetic field structure of post-flare loops after these loops have
had time to relax.  These loops have a similar appearance to the
reconnected loops very near the field reversal that are shown in
Figure \ref{paperplot_loops}.

Next we derive the loop-top positions along the inflow direction as a
function of height. By setting $B_y = 0$ and using Euler's formula and
trigonometric identities, we can obtain an equation for the height of
loop-tops as a function of $x$. The loop-top positions are given by
\begin{equation}
  \ylt(x) = \ln\left(\cos{a}\sec{x}\right). \label{looptop}
\end{equation}
For our purposes this expression is most useful near the field
reversal along the lower boundary at $(x,y)=(a,0)$ where the solution
is insensitive to the outer boundary.  We define $\theta =
\mathrm{arccot}\left( {\dif\ylt}/{\dif x} \right)$ as the (clockwise)
angle with respect to vertical for the loop-top positions.  Because
$\dif\ylt/\dif x = \tan{x}$ and $R\equiv B_L/B_R$, evaluating Equation
(\ref{looptop}) at $x=a$ yields
\begin{equation}
  \theta = 
  \frac{\pi}{2}
  \left(
    \frac{R-1}{R+1}
  \right).
  \label{magic}
\end{equation}
Equation (\ref{magic}) provides an upper limit on the distortion of
the asymmetric post-flare loops as a function of asymmetry.  The angle
from vertical for the loop-top positions is shown in both Figures
\ref{paperplot_loops} and \ref{analytic_solution}.  In Figure
\ref{paperplot_loops} we see that Equation (\ref{magic}) reliably
approximates the angle from vertical that the loop-top positions take
very near the field reversal.  Equation (\ref{magic}) overestimates
the angle slightly in the simulations because of the finite width of
the magnetic reversal along the lower boundary ($\delta_0=0.1$) and
because the field lines are not fully potential due in part to a
finite pressure gradient.

\section{OBSERVATIONAL SIGNATURES OF ASYMMETRIC RECONNECTION
  \label{observational}} 

In this section, we describe the observational signatures predicted by
our simulations of line-tied asymmetric reconnection during solar
eruptions.

\subsection{The Location of the X-line and Flow Reversal}

The location of the principal X-line is important in flare/CME current
sheets because it helps determine the partition of outflow energy
towards and away from the Sun.  These simulations suggest that, at
least during reconnection with a weak guide field, the principal
X-line will be located near the lower base of flare/CME current
sheets, not too far above the post-flare loops
\citep[e.g.,][]{murphy:retreat, shen:2011}.  The principal X-line and
flow reversal are probably separated by a distance that is shorter
than the observational errors and systematic uncertainties.  Finding
the location of the flow reversal in a current sheet viewed nearly
edge-on is difficult but can be done by tracking the motions of
current sheet blobs.  This task has been accomplished by
\citet{savage:2010}, who find that the flow reversal is at a height of
just {$\sim$}$0.25$ solar radii above the limb in the Cartwheel CME
current sheet.  For comparison, the post-flare loops have a height
between $0.1$ and $0.2$ solar radii above the limb and the current
sheet extends several solar radii outward into the LASCO field of
view.  Our simulations also predict that the X-line is located on the
weak magnetic field side of the current sheet (see Figure
\ref{paperplot_simulation}); however, we anticipate that this
signature is beyond our current observational capabilities.

\subsection{Post-Flare Loops and Loop Footpoints} 


A signature of line-tied asymmetric reconnection is the distortion or
skewing of post-flare loops as shown in Figure \ref{paperplot_loops}
and described in Section \ref{loops}.  The loop apexes are not
immediately above each other.  Rather, the apex positions along the
inflow direction are a function of height.  The post-flare loops
develop a characteristic skewed candle flame shape.  This signature
should be apparent in H$\alpha$, EUV, and X-ray observations of
post-flare loops during line-tied asymmetric reconnection.
A candidate event with a clear skewed candle flame shape is the 1992
February 21 flare on the east limb that was analyzed by
\citet{tsuneta:1992} and \citet{tsuneta:1996}.  However, projection
effects associated with a complicated three-dimensional geometry might
also lead to a skewed candle flame structure during some events
\citep[see, for example, Figure 15 of][]{forbes:acton:1996}.


Solar flares characteristically show hard X-ray (HXR) emission at the
footpoints of newly reconnected field lines in response to energetic
particles and the thermal conduction front impacting the chromosphere.
The standard model of solar flares predicts that the footpoints move
away from the neutral line so that the instantaneous location of the
footpoints is given by the amount of reconnected flux
\citep[e.g.,][]{lin:1995, forbes:acton:1996, lin:2004:ribbons}.  This
behavior has been observed during many flares
\citep[e.g.,][]{asai:2004, krucker:2005, yang:2009, yang:2011},
although more complicated motions are possible
\citep[e.g.,][]{bogachev:2005, sakao:1998, grigis:2005, somov:2005,
  ji:2006, su:2007}.  Because of the requirement that equal amounts of
flux be reconnected from each upstream region in two-dimensional
simulations, we predict that the velocity of the footpoint in the
strong field region will be slower than the velocity of the footpoint
in the weak field region.  For example, if the magnetic field in one
upstream region has twice the strength of the other region, then the
footpoints on the strong magnetic field side will move half as quickly
as the footpoints on the weak magnetic field side.
By combining observations during the 2003 October 29 X10 flare by the
\emph{Reuvan Ramaty High Energy Spectroscopic Imager} (\RHESSI) and
the Michelson Doppler Interferometer \citep[MDI,][]{scherrer:1995} on
\SOHO, \citet{krucker:2005} showed that the magnetic field near the
slower moving footpoint was generally stronger than in the faster
moving footpoint \citep[see also][]{svestka:1976}.  While this is
consistent with the predictions from our models, theory needs to take
into account the three-dimensional nature of magnetic reconnection and
the patchy distribution of magnetic flux in the photosphere.

Asymmetry in the post-flare loop structure will affect the relative
intensities from the emissions at each footpoint.  HXR emission is
largely determined by the transport of energetic particles from above
the loop-top into the chromosphere.  Energetic particles on the strong
magnetic field side will be more likely to be reflected because
magnetic mirroring is more effective.  Therefore, energetic particles
will be more likely to enter the lower solar atmosphere on the weak
magnetic field side.  Consequently, the 
footpoint on the weak magnetic field side is expected to yield
stronger HXR emission and chromospheric evaporation due to beam
heating \citep[][]{melrose:1979, sakao:1994, melrose:1981,
ghuang:2007}.
\citet{kundu:1995} present two flares observed by the Nobeyama
radioheliograph \citep{nakajima:1994} and the Hard X-ray Telescope on
\Yohkoh\ \citep{kosugi:1991} in support of this scenario.
The footpoints with weak HXR emission (indicating fewer particles
escaping from the trap) had stronger gyrosynchrotron emission
(suggestive of a stronger magnetic field).
While this behavior does not happen in all flares \citep{goff:2004},
additional sources of asymmetry include an asymmetric initial pitch
angle distribution \citep[e.g.,][]{ghuang:2010}, differences in the
column density in each footpoint \citep[see,
however,][]{sainthilaire:2008}, and directionality in the accelerating
electric field \citep{hamilton:2005, li:lin:2012}.

\subsection{Reconnection Inflow Velocities}

Several works have reported observations of reconnection inflow
velocities associated with flare/CME current sheets
\citep[e.g.,][]{yokoyama:2001, lin:2005, narukage:2006, takasao:2012}.  During
asymmetric inflow reconnection, the inflow velocities on either side
of the current sheet are predicted to differ.  In a steady-state, one
would expect from flux conservation that the out-of-plane electric
field will be constant, leading to the relation
\begin{equation}
  V_LB_L = V_RB_R.\label{spoo}
\end{equation}
This implies that the ratio of inflow velocities is inversely
proportional to the ratio of upstream magnetic field strengths and
that in principle measuring the ratio of inflow velocities on either
side of the current sheet would directly provide the upstream magnetic
field ratio.  

Figure \ref{paperplot_inflow} shows that there are strong variations
in the electric field in the simulation reference frame and that the
inflow component of velocity is approximately zero in the strong
upstream region.  Thus the reconnection process in the simulations is
not time-independent.  Therefore, Equation (\ref{spoo}) should not be
expected to give reliable estimates of the magnetic field asymmetry
except by taking the velocities in the reference frame of the X-line
and showing that the reconnection process is steady.  However, despite
the inapplicability of Equation (\ref{spoo}) during time-dependent
asymmetric reconnection, a systematic demonstration that the
reconnection inflows differ on either side of the current sheet will
provide suggestive evidence in future observations that the
reconnection process is asymmetric.


\subsection{Drifting of the Current Sheet}

A commonly observed feature of CME current sheets is that they appear
to drift or tilt with time.  \Hinode/XRT observations of the Cartwheel
CME show that the current sheet drifted at a rate of
{$\sim$}$4^\circ$~hr$^{-1}$ \citep{savage:2010}.  \citet{ko:2003}
discussed observations on 2002 January 8 of a current sheet with a
drift of {$\sim$}{$0.8^\circ$}~hr$^{-1}$.  These drift rates are
significantly greater than can be accounted for by solar rotation.

There are several possible explanations for the observed drifting of
CME current sheets:
(1) \citet{savage:2010} suggest that the Cartwheel CME current sheet
is observed at an angle to the plane-of-sky and that the appearance of
drifting is caused by different parts of the current sheet actively
reconnecting at different times.
(2) The tilting could be due to the drifting of the X-line and current
sheet into the strong magnetic field region during line-tied
asymmetric reconnection as discussed in this paper.  The predicted
velocity of less than a percent of the Alfv\'en speed is within
observational constraints.  However, the simulations show drifting but
not the observed tilting of the current sheet. 
(3) The rising flux rope could pull the plasma sheet region along with
it so that the drifting is caused by macroscopic behavior.  This
mechanism requires that the current sheet become more aligned with the
direction of flux rope propagation.  For the Cartwheel CME, however,
this appears to not be the case.
(4) There is a large-scale force imbalance between the two upstream
regions so that the entire region surrounding the current sheet is
pushed rapidly towards one direction.  Such a force imbalance could
easily occur in the early stages of an ejection
\citep[see][]{ko:2003}.
(5) The tilting is caused by relaxation in the post-eruption active
region as the magnetic field configuration becomes more potential.

Further numerical and observational tests are required to constrain
which mechanisms lead to current sheet drifting during CMEs\@.  The
location of the CME current sheet relative to the direction of
propagation of the flux rope may be of particular importance in
determining how upflow from the current sheet impacts and influences
the evolution of the rising plasmoid.  However, it is not yet known if
CME current sheets are an important component of the energy and
momentum budgets of CMEs\@.

\subsection{Circulation Within the Rising Flux Rope}

In Section \ref{plasmoid} we show that the asymmetric inflow condition
allows net vorticity to develop in the outflow plasmoid.  This
circulation pattern develops because the outflow jet impacts the
rising flux rope obliquely rather than directly at its base.  When the
current sheet and direction of flux rope propagation are misaligned,
this leads to the possibility that the reconnection outflow jet
torques the rising plasmoid.  Vortex motions like those seen in Figure
\ref{paperplot_vorticity} are not unique to configurations with a
line-tied lower boundary, but rather should be a generic feature of
reconnection with asymmetric upstream magnetic fields such as at the
dayside magnetopause.

\citet{martin:2003}, \citet{panasenco:2008}, and
\citet{panasenco:2011} report observations of several CMEs that
display a rolling motion about the axis of the erupting prominence.
We hypothesize that this rolling motion is induced by an offset
between the CME current sheet and the rising flux rope during some
events.  However, it is unknown if the kinetic energy released by
reconnection is enough to drive this circulation.

There are other candidate mechanisms for the development of apparent
circulation and vorticity in the rising flux rope when viewed in
cross-section.  A similar flow pattern could develop by untwisting of
the magnetic field in the rising flux rope during expansion and
relaxation.  Alternatively, the external magnetic field could deflect
the flux rope's outward motion, thus leading to apparent rolling
behavior.  For example, \citet{panasenco:2011} note the presence of
coronal holes near CMEs that display rolling motion.  This rolling
behavior about the flux rope axis is in contrast to rotation about the
direction of CME propagation \citep[e.g.,][]{lynch:2009,
patsourakos:2011, thompson:2012}.

\subsection{Ultraviolet Spectroscopy of the Inflow Regions and Current
  Sheet}

If there is a strong asymmetry in the plasma conditions on the two
sides of the current sheet, an asymmetry in the emission line
intensities might be expected as well.  An example is seen in Figure
12 of \citet{ko:2003}, where the [\ion{Fe}{18}] and [\ion{Ca}{14}]
lines are sharply peaked at the same position along the slit, while
the \ion{Si}{12} emission is shifted about $70''$ (one bin along the
entrance slit, or $50$ Mm) toward the north, and [\ion{Fe}{12}] is
shifted still farther north by another $70''$.  Lower temperature
lines such as [\ion{Fe}{10}] are entirely absent from the immediate
area of the current sheet, but a definite [\ion{Fe}{10}] feature is
seen about $140''$ to the south.

There is a clear asymmetry in the temperature between the north and
south sides of the current sheet in Figure 12 of \citet{ko:2003}.  It
is hard to estimate the temperatures accurately because of the
ambiguity caused by the uncertainty in the foreground and background
contributions to the line intensities.  Roughly speaking, the
[\ion{Fe}{10}] and [\ion{Fe}{12}] intensities on the southern side are
comparable, indicating a temperature $\log{T} \sim 5.95$, while on the
northern side \ion{Si}{12} is several times stronger than
[\ion{Fe}{12}], suggesting $\log{T} \sim 6.15$.  

The density contrast is more difficult to estimate.  Based on those
temperatures, the emission measure on the northern side is four times
that to the south.  Assuming that the depth along the line of sight is
the same on both sides of the current sheet, the density is twice as
high to the north.  Another density diagnostic is the \lyb\ to \lya\
ratio.  Since \lyb\ contains comparable contributions from
collisionally excited ($\propto n^2$) and radiative scattering
($\propto n$) components, while \lya\ is almost entirely formed by
radiative scattering, the ratio is a reliable diagnostic.
Unfortunately, the separation of the emission from the region of the
current sheet from broad background emission is even more difficult
than it is for the {Fe} and {Si} lines discussed above.  We estimate
that the \lyb\ to \lya\ ratios are $0.01$ and $0.006$ on the southern
and northern sides, respectively, which would imply that the density
is $\sim${$1.5$} times higher in the south than in the north. Each of
these density diagnostics is accurate to within about a factor of two,
suggesting that the northern and southern densities are comparable to
within uncertainties.  Overall, we conclude that the plasma pressure
is somewhat higher in the north than the south.  It would be dangerous
to conclude that the magnetic pressure is higher in the south than the
north, however, since the region is probably not in equilibrium.

Asymmetry in the upstream temperatures might also affect the charge
state distribution in the current sheet plasma and the rising flux
rope.  For example, the two upstream regions could start in ionization
equilibrium at $T_L = 1$ and $T_R = 2$ MK\@.  Plasma entering the
current sheet will be heated very quickly to {$\sim$}$5$--$10$ MK\@.
The ionization time scales typically range from $10$--$1000$ s and are
a function of both temperature and density.  This is comparable to a
dynamical time scale and thus we expect the plasma leaving the current
sheet to be underionized \citep{ko:2010, imada:2011:nei}.  Because the
current sheet plasma contains contributions from upstream plasma with
different initial temperatures, this may result in a broader charge
state distribution than would be expected if there was just one
starting temperature.  However, the difference may be small because
the time scales to ionize low charge state species will be short in
very hot plasma.
Differences in the inflow velocities, densities, and temperatures
might also lead to observable differences in the non-thermal line
widths between the two upstream regions.

\section{DISCUSSION AND CONCLUSIONS \label{discussion}}

In this article, we present resistive MHD simulations of line-tied
asymmetric reconnection in the context of solar flare and CME current
sheets.  There is asymmetry along the inflow direction because of
different upstream densities and magnetic field strengths, and
asymmetry along the outflow direction because the initial perturbation
is placed near a conducting wall.  The simulations are used to
understand the basic physics of asymmetric reconnection and predict
observational signatures of asymmetric reconnection in the solar
atmosphere.  This approach provides a unified picture of how asymmetry
in the reconnection process impacts flare emissions and CME evolution.

As in previous simulations of asymmetric outflow reconnection, the
X-line is generally located near the lower base of the current sheet.
Consequently, the outflow velocity in the unobstructed direction is
significantly faster than the outflow velocity in the obstructed
direction.  The slow downflows occur in simulations for two reasons.
First, the downflows impact a region of strong magnetic and plasma
pressure, thus slowing them down.  Second, the principal X-line is
often located near the lower base of the current sheet so that the
downward tension force is much weaker compared to the upward tension
force.  Downflowing loops in flare/CME current sheets are frequently
observed to propagate at velocities several times slower than the
Alfv\'en speed \citep{mckenzie:1999, asai:2004, sheeley:2004,
  reeves:2008:shrinkage, savage:2011, warren:2011, mckenzie:2011}, in
contrast to symmetric models of reconnection that predict
bidirectional Alfv\'enic jets.  Simulations such as those presented in
this paper suggest that the outflow asymmetry is responsible for this
difference in outflow speeds \citep[see also][]{roussev:2001A,
  galsgaard:2002, reeves:2010, murphy:retreat, shen:2011}.
Observational analysis techniques such as those by \citet{savage:2010}
that constrain the position of the reconnection flow reversal are
important for resolving the problem of slow downflows.  We do note
that there are observations of fast downflows during some reconnection
events \citep[e.g.,][]{innes:2003B}.

In simulations with asymmetric upstream magnetic fields, we find that
the post-flare loops are skewed when compared to cases with symmetric
inflow.  The loop-tops are not directly above each other; rather, the
positions of the loop-tops are a function of height.  The observed
shape of these loops is reminiscent of a candle flame.  The structure
of near-potential loops close to the field reversal along the lower
boundary are approximately given by our analytic solution in Section
\ref{analytic}.  Cuspy post-flare loops with an apparent candle flame
structure have been observed during many events
\citep[][]{tsuneta:1992, tsuneta:1996, forbes:acton:1996,
  reeves:2008:shrinkage}.  Observations showing such behavior would
allow us to diagnose or place limits on the asymmetry in the upstream
magnetic fields.  However, 3D geometry and projection effects would
need to be considered carefully because they could also contribute to
a skewed post-flare loop appearance \citep[see][]{forbes:acton:1996}.

The morphological features associated with line-tied asymmetric
reconnection grow preferentially into the weak magnetic field upstream
region.  Unlike in cases with symmetric inflow, there is net vorticity
in the outflow plasmoid because the reconnection outflow jet impacts
it obliquely.  This result suggests that CME current sheets could
drive a circulation pattern in erupting flux ropes when viewed in
cross-section for events where the current sheet is offset from the
direction of flux rope motion.

The X-line retreats away from the obstructing wall in all simulations
where it is present \citep[see also][]{oka:2008, murphy:retreat},
while generally staying near the lower base of the current sheet.  The
exception is case F with a strong guide field which shows that the
X-line becomes located near the top exit of the current sheet.  The
X-line drifts towards the upstream region with the stronger magnetic
field in most simulations with asymmetric inflow
\citep[e.g.,][]{ugai:2000, cassak:asym}.  During simulations with
symmetric upstream magnetic fields but asymmetric upstream densities,
the X-line does not drift significantly along the inflow direction.

For each simulation we find that the plasma flow velocity at the
X-line is substantially different from the time derivative of the
X-line's position.  Any difference between these two velocities must
be due to resistive diffusion \citep[e.g.,][]{seaton:2008,
murphy:retreat}.  With the exception of the guide field and symmetric
outflow cases, the X-line retreats along the outflow direction against
the flow of the plasma.  This occurs because diffusion of the inflow
component of the magnetic field along the inflow direction is able to
shift the X-line position \citep{murphy:retreat}.  During reconnection
with asymmetric inflow, the inflow component of plasma velocity at the
X-line is in the same direction as the time derivative of the X-line
position along the inflow direction.  However, the plasma flow
velocity is much faster than the rate of change in position of the
X-line along the inflow direction.  Consequently, one must be careful
when converting simulations to the reference frame of the X-line
because the plasma flow velocity will likely be different.

While our simulations provide significant insight into the basic
physics and observational consequences of line-tied asymmetric
reconnection in the solar atmosphere, it is important to note the
limitations of our models.  Most noticeably, $\beta$ is larger in the
simulations than in most of the corona.  Because the simulations start
from a perturbed initial equilibrium, total pressure balance is
enforced along the inflow direction by increasing plasma pressure on
the weak magnetic field side.  It is difficult to maintain a large
asymmetry in low-$\beta$ plasmas.  The asymmetries in the simulations
are probably more extreme than in the solar atmosphere.  We speculate
that magnetic field asymmetries of {$\sim$}$10$--$25$\% are common,
with differences of perhaps {$\gtrsim$}$50$\% possible in more extreme
cases.  The assumption of an initial equilibrium is also questionable
since CMEs are non-equilibrium events.  The simulations do not include
radiative cooling, coronal heating (except for viscous and Ohmic
heating), or initial vertical stratification of the solar atmosphere.
The outer conducting wall boundary conditions will affect the results
on long time scales by additional line-tying as well as pileup of
reconnection exhaust.  The Lundquist numbers in our simulations are
{$\lesssim$}$10^4$ so our current sheets are below the $S\sim 5\times
10^4$ threshold for the onset of the plasmoid instability
\citep[e.g.,][]{loureiro:2007, bhattacharjee:2009, huang:2010,
shepherd:2010, barta:2008, barta:2011A, ni:2010, shen:2011,
biskamp:1986}.  Our simulations consequently have only one X-line.

Future work on this problem should be performed using a combination of
observations and improved numerical simulations.  In particular, more
realistic initial configurations will provide more detailed
predictions that can be compared directly to observations.  The HyLoop
suite of codes \citep{winter:thesis, winter:2011} has the ability to
take asymmetric loop configurations directly from these simulations,
inject a population of energetic particles, and predict flare
emissions in detail.  Observationally, investigations of candle flame
post-flare loops, current sheet drifting, and footpoint motion and
relative intensities will provide constraints on the theoretical
models.  In particular, events that display multiple signatures will
provide the most useful constraints and provide the most complete
story.

Many open questions remain for asymmetric reconnection in the solar
atmosphere.  They include: 
(1) How asymmetric are typical flare/CME current sheets? 
(2) What mechanisms are responsible for asymmetric HXR footpoint
    emission during flares?
(3) What causes the drifting motion observed in many flare/CME current
sheets?
(4) Where is the principal X-line in these current sheets?
(5) How important are CME current sheets to the eruption as a whole?
(6) Are CME current sheets able to instigate circulation in the rising
flux rope?  
(7) What are the effects of the patchy distribution of magnetic
flux in the photosphere?  
We hope to address these problems in future work.

\acknowledgements

The authors thank 
  P.~A.\ Cassak, 
  S.~E.\ Guidoni,
  Y.-K.\ Ko, 
  D.~E.\ McKenzie,
  L.\ Ni,
  M.\ Oka,
  S.~L.\ Savage,
  C.\ Shen, 
  C.~R.\ Sovinec, 
  H.~P.\ Warren,
  D.\ Webb,
  Y.-H.\ Yang,
  and S.\ Zenitani 
for useful discussions.
This research is supported by NASA grants NNX09AB17G and NNX11AB61G
and contract NNM07AB07C to the Smithsonian Astrophysical Observatory.
M.P.M.\ acknowledges support from several NASA grant NNX09AH22G.
C.L.P.\ acknowledges support from the NSF-REU solar physics program at
the Center for Astrophysics, grant number ATM-0851866.  
K.K.R.\ is supported under the NSF-SHINE program, grant number
ATM0752257\@.
D.B.S.\ acknowledges support from the Belgian Federal Science Policy
Office through ESA-PRODEX grant number 4000103240\@.
The work of J.L.\ was also supported by the Program 973 grant
2011CB811403, the NSFC grant 10873030, and the CAS grant KJCX2-EW-T07
to the Yunnan Astronomical Observatory.
N.A.M.\ acknowledges the hospitality of the Yunnan Astronomical
Observatory during a visit supported by CAS grant 2010Y2JB16\@.
The authors thank members of the NIMROD team for code development
efforts that helped make this work possible.
Resources supporting this work were provided by the NASA High-End
Computing (HEC) Program through the NASA Advanced Supercomputing (NAS)
Division at Ames Research Center.
This article has benefited greatly from the use of NASA's Astrophysics
Data System.



\begin{deluxetable}{ccccccccl}
\tabletypesize{\scriptsize}
\tablecaption{Simulation Parameters}
\tablewidth{0pt}
\tablehead{
\colhead{Case} &
\colhead{$m_x$} &
\colhead{$m_y$} &
\colhead{$V_{Ah0}$} &
\colhead{$R_0$} &
\colhead{$f_0$} &
\colhead{$B_{z0}$} &
\colhead{$S_h$\tablenotemark{a}} &
\colhead{Notes}
}
\startdata
A & 24 & 80 & 1.00 & $1.0$   & $1$ & $0$ & $8000$ & Symmetric inflow\\
B & 60 & 80 & 0.71 & $0.5$   & $1$ & $0$ & $4000$ & Asymmetry in the upstream magnetic field\\
C & 60 & 80 & 0.50 & $0.25$  & $1$ & $0$ & $2000$ & Asymmetry in the upstream magnetic field\\
D & 60 & 80 & 0.35 & $0.125$ & $1$ & $0$ & $1000$ & Asymmetry in the upstream magnetic field\\
E & 60 & 80 & 0.50 & $0.25$  & $1$ & $0$ & $667$  & Like case C but with triple resistivity\\
F & 60 & 80 & 0.50 & $0.25$  & $1$ & $4$ & $2000$ & Like case C but with a guide field\\
G & 32 & 84 & 0.63 & $1.0$   & $4$ & $0$ & $5040$ & Symmetric magnetic field, asymmetric density\\
H & 60 & 48 & 0.50 & $0.25$  & $1$ & $0$ & $8000$ & Symmetric outflow, periodic in $y$-direction
\enddata 
\tablenotetext{a}{Using Equation (\ref{lundquist}) assuming $L=8$.}
\label{simtable}
\end{deluxetable}

\clearpage



\end{document}